\documentclass[12pt]{article}
\usepackage{amstex,amssymb,cite}

\begin{document}
\begin{flushright}
SU-4240-688 \\
TIFR/TH/98-44 \\
\end{flushright}

\begin{center}
{\large{\bf MONOPOLES AND SOLITONS IN FUZZY PHYSICS}}

\bigskip 
S. Baez$^a$, A. P. Balachandran$^a$, S. Vaidya$^b$ and B. Ydri$^a$ \\
$^a${\it Physics Department, Syracuse University, \\
Syracuse,N.Y.,13244-1130, U.S.A.}

$^b${\it Tata Institute of Fundamental Research, \\
Colaba, Mumbai, 400 005, India.}

\end{center}

\begin{abstract}  
Monopoles and solitons have important topological aspects like
quantized fluxes, winding numbers and curved target spaces. Naive
discretizations which substitute a lattice of points for the
underlying manifolds are incapable of retaining these features in a
precise way. We study these problems of discrete physics and matrix
models and discuss mathematically coherent discretizations of
monopoles and solitons using fuzzy physics and noncommutative
geometry. A fuzzy ${\sigma}$-model action for the two-sphere
fulfilling a fuzzy Belavin-Polyakov bound is also put forth.
\end{abstract}

\newpage
A fuzzy space
(\cite{madore,gropre,grklpr3,grklpr1,grklpr2,watamura1,watamura2,frgrre})
is obtained by quantizing a manifold, treating it as a phase space. An
example is the fuzzy two-sphere $S^2_{F}$. It is described by
operators $x_{i}$ subject to the relations $\sum_{i} x_i^2 =1$ and
$[x_{i},x_{j}]=(i/{\sqrt{l(l+1)}}) {\epsilon}_{ijk}x_{k}$. Thus
$L_i=\sqrt{l (l+1)}x_i$ are $(2l+1)$-dimensional angular momentum
operators while the canonical classical two-sphere $S^2$ is recovered
for $l{\rightarrow}{\infty}$. Planck's work shows that quantization
creates a short distance cut-off, therefore quantum field theories
(QFT's) on fuzzy spaces are ultraviolet finite. If the classical
manifold is compact, it gets described by a finite-dimensional matrix
model, the total number of states being finite too. Noncommutative
geometry (\cite{madore,connes,landi,coquereaux,vargra,mssv,varilly})
has an orderly prescription for formulating QFT's on fuzzy spaces so
that these spaces indeed show us an original approach to discrete
physics.

In this note, we focus attention on $S^2_{F}$ and discuss certain of
its remarkable aspects, entirely absent in naive discretizations.
Quantum physics on $S^2_{F}$ is a mere matrix model, all the same they
can coherently describe twisted topologies like those of monopoles and
solitons. Traditional attempts in this direction based on naive
discrete physics have at best been awkward having ignored the
necessary mathematical structures (projective modules and cyclic
cohomology). Not all our results are new, our construction of
monopoles being a reformulation of the earlier important work of
Grosse et al. \cite{grklpr1}.

\section{\it Classical Monopoles and $\sigma$-Models}
There is an algebraic formulation of monopoles and solitons suitable
for adaptation to fuzzy spaces. We first outline it using the case of
$S^2$(\cite{mssv}).

Let ${\cal A}$ be the commutative algebra of smooth functions on
$S^2$. Vector bundles on $S^2$ can be described by projectors ${\cal
P}$. ${\cal P}$ is a matrix with coefficients in ${\cal A}({\cal
P}_{ij}{\in}{\cal A}),$ and fulfills ${\cal P}^2={\cal P}$ and ${\cal
P}^{\dagger}={\cal P}$. If the points of $S^2$ are described by unit
vectors $\vec{n} {\in} {\mathbb R}^3$, the projector for unit monopole
charge is ${\cal P}^{(1)}=(1 + \vec{\tau}.\hat{n})/2$ where ${\tau}_i$
are Pauli matrices and $\hat{n}_i$ are coordinate functions,
$\hat{n}_i (\vec{n})=n_i$. If ${\cal A}^{2^{N}} = {\cal
A}{\otimes}{\mathbb C}^{2^{N}}$ consists of $2^{N}$-component vectors
${\xi}=({\xi}_1, {\xi}_2,..., {\xi}_{2^{N}})$, ${\xi}_i{\in} {\cal
A}$, then the sections of vector bundles for monopole charge 1 are
${\cal P}^{(1)}{\cal A}^{2}$. For monopole charge $\pm N $ $ (N>0)$,
the corresponding projectors are ${\cal P}^{(\pm N)}=\prod
_{i=1}^{i=N}(1 \pm \vec{\tau}^{(i)}.\hat{n})/2$ where
$\vec{\tau}^{(i)}$ are commuting sets of Pauli matrices. They give
sections of vector bundles ${\cal P}^{(\pm N)}{\cal A}^{2^{N}}$ with
$\vec{\tau}^{(i)}.\hat{n}\;{\cal P}^{(\pm N)}{\xi}=\pm {\cal P}^{(\pm
N)}{\xi}$, $\vec{\tau}^{(i)}$ acting on the $i^{th}$ ${\mathbb C}^2$
factor. For the trivial bundle, we can use ${\cal P}^{0}=(1 +
{\tau}_3)/2$ (or $(1 - {\tau}_3)/2$), or more simply just the
identity.

The self-same projectors also describe nonlinear $\sigma$-models. To
see this, consider the projector ${\cal P}^{(0)}$ and its orbit
$\langle h{\cal P}^{(0)}h^{-1} : h{\in}SU(2) \rangle = S^{2}$. If now
we substitute for $h$ a field $g$ on $S^{2}$ with values in $SU(2)$,
each $g{\cal P}^{(0)}g^{-1}$ describes a map $S^2 {\rightarrow}
S^2$. It is a $\sigma$-model field on $S^2$ with target space $S^2$
and zero winding number. (Winding number is zero as $g$ can be
deformed to a constant map). For winding number 1, it is appropriate
to consider the orbit of ${\cal P}^{(1)}$ under $g$. For fixed
$\vec{n}$, as $g(\vec{n})$ is varied, $g(\vec{n}) \frac{ 1 +
\vec{\sigma}.\hat{n}(\vec{n})}{2} g(\vec{n})^{-1}$ is still $S^2$ so
that as $\vec{n}$ is varied, we get a map $S^2{\rightarrow}S^2$. (More
correctly we get the section of an $S^2$ bundle over $S^2$). For
winding number $\pm{N}$, we can consider the orbit of ${\cal P}^{(\pm
N)}$ under conjugation by $g^{\tilde{\otimes}N}$'s where
$g^{\tilde{\otimes}N} (\vec{n})=
g(\vec{n}){\otimes}g(\vec{n}){\otimes} \cdots {\otimes}g(\vec{n})$
($N$ factors). Here the $i$th $g(\vec{n})$ acts only on
$\vec{\tau}^{(i)}$.

\section{\it Winding Numbers for the Classical Sphere} 
What about formulas for invariants like Chern character and winding
number? The ideal way is to follow Connes
(\cite{connes,landi,vargra,varilly,coquereaux,mssv}) and introduce the
Dirac and chirality operators
\begin{eqnarray}
{\cal D}&=&{\epsilon}_{ijk} {\sigma}_{i} \hat{n}_j {\cal J}_{k},
\nonumber\\ 
{\Gamma}&=&{\sigma}.\hat{n} \label{contoprs}
\end{eqnarray}
where ${\sigma}_i$ are Pauli matrices, $\vec{\cal
J}=-i(\vec{r}{\times}{\vec{\bigtriangledown}})+\frac{\vec{\sigma}}{2}$
is the total angular momentum and $\hat{n}=\vec{r}/|\vec{r}|$. The
important points to keep in mind here are the following:

i){\it ${\Gamma}$ commutes with elements of ${\cal A}$ and
anti-commutes with ${\cal D}$.}

ii){\it ${\Gamma}^2 = 1$ and ${\Gamma}^{\dagger}={\Gamma}$.}

The Chern numbers (or the quantized fluxes) for monopoles then are 
\begin{equation}
\pm N= \frac{1}{4 \pi}\int d (\cos{\theta})d{\phi}\;{\rm
Tr}\;{\Gamma}{\cal P}^{(\pm N)}\;[{\cal D}, {\cal P}^{(\pm N)}]\;[{\cal
D}, {\cal P}^{(\pm N)}](\vec{n}).
\label{wno}
\end{equation}
They do not change if ${\cal P}^{(\pm N)}$ are conjugated by
$g^{\tilde{\otimes}N}$ and are also the soliton winding numbers.

\section{\it Fuzzy Monopoles}
The algebra $A$ generated by $x_i$ is the full matrix algebra of
$(2l+1){\times}(2l+1)$ matrices. Fuzzy monopoles are described by
projectors $p^{({\pm} N)}\; (p^{({\pm} N)}_{ij} {\in} A)$, which as
$l{\rightarrow}{\infty}$ approach ${\cal P}^{({\pm} N)}$. We can find
them as follows: For $N=1$, we can try $(1 + \vec{\tau} . \vec{x})/2$,
but that is not an idempotent as the $x$'s do not commute. We can fix
that though: since $(\vec{\tau}.\vec{L})^2 = l(l+1) -
\vec{\tau}.\vec{L}$, ${\gamma}_{\tau}=\frac{1}{l+1/2}
(\vec{\tau}.\vec{L}+ 1/2)$ squares to 1 as first remarked by Watamuras
(\cite{watamura1,watamura2}). Hence $p^{(1)}=(1 + {\gamma}_{\tau})/2$.

There is a simple interpretation of $p^{(1)}$. We can combine
$\vec{L}$ and ${\vec{\tau}}/2$ into the $SU(2)$ generator
${\vec{K}^{(1)}}=\vec{L} + {\vec{\tau}}/2$ with spectrum $k(k+1)$
(with $k=l \pm1/2$) for
${K^{(1)}}^2{\equiv}{\vec{K}}^{(1)}.{\vec{K}}^{(1)}$. The projector to
the space with the maximum $k$, namely $\frac{{K^{(1)}}^{2} -
(l-1/2)(l+1/2) }{(l+1/2)(l+3/2)-(l-1/2)(l+1/2)}$, is just $p^{(1)}$.

This last remark shows the way to fuzzify ${\cal P}^{(N)}$. We
substitute $\vec{K}^{(N)}=\vec{L} + \sum_{i=1}^{i=N}
(\vec{\tau}^{(i)})/2$ for $\vec{K}^{(1)}$ and consider the subspace
where ${K^{(N)}}^{2}{\equiv}{\vec{K}}^{(N)}.{\vec{K}}^{(N)}$ has the
maximum eigenvalue $k_{max}(k_{max}+1),k_{max}=l+N/2$. On this space
$(\vec{L} + \vec{\tau}^{(i)}/2)^2$ has the maximum value
$(l+1/2)(l+3/2)$ and $\vec{\tau}^{(i)}.\vec{L}$ is hence $l$. Since
$\vec{\tau}^{(i)}.\vec{x}$ approaches $\vec{\tau}^{(i)}.\hat{n}$ on
this subspace as $l \rightarrow \infty$, $p^{(N)}$ is just its
projector:
\begin{equation}
p^{(N)}\equiv p^{(+N)}={\frac{{\prod}_{k{\neq}k_{max}} [{K^{(N)}}^{2}-
k(k+1) ]}{{\prod}_{k{\neq}k_{max}} [ k_{max}(k_{max}+1) - k(k+1) ]}}.  
\end{equation}
$p^{(-N)}$ comes similarly from the least value $k_{min}=l-N/2$ of
$k$. [We assume that $2l \geq N$.]

We remark that the limits as $l{\rightarrow}{\infty}$ of $p^{(\pm N)}$
are exactly ${\cal P}^{(\pm N)}$, and not say ${\cal P}^{(\pm N)}$
times another projector. That is because if $\vec{\tau}^{(i)}.\vec{L}$
are all $l$, then $\vec{\tau}^{(i)}.\vec{\tau}^{(j)}={\bf 1}$ for all
$i{\neq}j$ and hence $k_{max}=l + \frac{N}{2}$. A proof goes as
follows. Vectors with ${\vec{L}}^{2}=l(l+1){\bf 1}$ can be represented
as symmetric tensor products of $2l$ spinors, with components
$T_{a_1...a_{2l}}$. The vectors with $\vec{\tau}^{(i)}.\vec{L}=l{\bf
1}$ as well have components $T_{a_1...a_{2l},b_1...b_N}$ with symmetry
under exchange of any $a_i$ with $a_j$ or $b_k$. So they are symmetric
under all exchanges of $b_i$ and $b_j$ and have
$(\frac{\vec{\tau}^{(i)}}{2} +\frac{\vec{\tau}^{(j)}}{2} )^2=2$,
$\vec{\tau}^{(i)}.\vec{\tau}^{(j)}={\bf 1}$.

Having obtained $p^{(\pm N)}$, we can also write down the analogues of
${\cal P}^{(\pm N)}{\cal A}^{2^{N}}$: they are the ``projective
modules'' $p^{(\pm N)}A^{2^{N}}, A^{2^{N}}= \langle
(a_1,a_2,..,a_{2^{N}}):a_i{\in} A\rangle$, and are the noncommutative
substitutes for sections of vector bundles.

If $(a_1,a_2,...,a_{2^N}) $ is regarded as a column, then column
dimension of $p^{(\pm N)}A^{2^{N}}$ is $L+1 {\equiv}
2(l{\pm}{\frac{N}{2}})+1$ and its row dimension is $M + 1 {\equiv}
2l+1$. Their difference is $\pm N$. This means that $p^{(\pm
N)}A^{2^{N}}$ can be identified with ${\hat{\cal H}}_{L,M}$ of
ref.\cite{grklpr1} where of course $L - M ={\pm} N$. In particular
angular momentum acts on $p^{(\pm N)}A^{2^{N}}$ via $\vec{K}^{(N)}$ on
left (they commute with $p^{(\pm N)}$) and $-\vec{L}$ on right, while
there are similar actions of angular momentum on ${\hat{\cal
H}}_{L,M}$ [see ref. \cite{grklpr1}].

\section{\it Fuzzy ${\sigma}$-models}
In the fuzzy versions of $\sigma$-models on $S^2$ with target $S^2$,
$g$ becomes a $2{\times}2$ unitary matrix $u$ with $u_{ij}{\in} A$.
Therefore $u{\in}U(2(2l+1))$. (We can impose $\det u=1$, that makes no
difference). An appropriate generalization $ u^{\tilde{\otimes}N} $ of
$g^{\tilde{\otimes}N}$ can be constructed as follows. If $C$ and $D$
are $2{\times}2$ matrices with entries $C_{ij}$, $D_{ij}{\in}A$, we
can define $Ca$ and $aD$ for $a \in A$ by $(Ca)_{ij}=C_{ij}a$ and
$(aD)_{ij} = aD_{ij}$. Let $C{\otimes}_{A}D$ denote the tensor product
of $C$ and $D$ over $A$ where by definition $Ca{\otimes}_{A}D =
C{\otimes}_{A}aD$. This definition can be extended to more
factors. For example, $C{\otimes}_{A}D{\otimes}_{A}E$ has the
properties $Ca{\otimes}_{A}D{\otimes}_{A}E =
C{\otimes}_{A}aD{\otimes}_{A}E$, $C{\otimes}_{A}Da{\otimes}_{A}E =
C{\otimes}_{A}D{\otimes}_{A}aE$. Then:
\begin{equation}
u^{\tilde{\otimes}N}=u{\otimes}_{A}u{\otimes}_{A} \ldots
{\otimes}_{A}u ~(~N \rm{~factors}).
\end{equation}
We can understand this construction in familiar terms by writing
$u={\bf 1}_{2{\otimes}2}a_{0}$ $+ {\tau}_{j}a_{j} =
{\tau}_{\mu}a_{\mu}(a_{\mu}{\in}A)$ where ${\tau}_0={\bf
1}_{2{\otimes}2}$. $[$Greek subscripts run from $0$ to $3$, Roman ones
from $1$ to $3]$. Unitarity requires that
\begin{eqnarray}
{\tau}_{\mu}{\tau}_{\nu}a_{\mu}^{*}a_{\nu}&=&{\bf 1},\nonumber\\
a_{\mu}^{*}& \equiv& a_{\mu}^{\dagger}. \label{unitarya}
\end{eqnarray}
In this notation,
$u{\otimes}_{A}u={\tau}_{\mu}{\otimes}{\tau}_{\nu}a_{\mu}a_{\nu}$,
${\otimes}({\equiv}{\otimes}_{\mathbb C})$ denoting Kronecker
product. It is also
${{\tau}_{\mu}}^{(1)}a_{\mu}{{\tau}_{\nu}}^{(2)}a_{\nu}$ in an evident
notation. Proceeding in this way, we find,
\begin{equation}
u{\otimes}_{A}u{\otimes}_{A} \ldots {\otimes}_{A}u =
{\tau}^{(1)}_{{\mu}_1}a_{{\mu}_1} {\tau}^{(2)}_{{\mu}_2}a_{{\mu}_2}
\ldots {\tau}^{(N)}_{{\mu}_N}a_{{\mu}_N}.
\end{equation}
It is unitary in view of (\ref{unitarya}).

The significant point here is that $u^{\tilde{\otimes}N}$ is a matrix
with coefficients in $A$ and not $A{\otimes}A{\otimes} \ldots
{\otimes}A$ as is the case for $u{\otimes}u{\otimes}u \ldots
{\otimes}u$. We remark that $g^{\tilde{\otimes}N}$ can also be written
as $g{\otimes}_{\cal A}g{\otimes}_{\cal A}g \ldots {\otimes}_{\cal
A}g$ ($N$ factors). It is then a function only of $\vec{n}$ and has
the meaning stated earlier.
  
The orbits of $p^{(\pm N)}$ under conjugation by
$u^{\tilde{\otimes}N}$ are fuzzy matrix versions of $\sigma$-model
fields with winding numbers $\pm{N}$. [Here we take $p^{(0)}$ to be
$(1+\tau_3)/2$ say and its $u^{\tilde{\otimes}N}$ to be $u$
itself. Henceforth our attention will be focused on $N \neq 0$.]

\section{\it ``Winding Numbers'' for the Fuzzy Sphere}
The fuzzy Dirac operator $D$ and chirality operator ${\gamma}$ are
important for writing formulae for the invariants of projectors. There
are proposals for $D$ and ${\gamma}$ in
\cite{gropre,grklpr1,grklpr2,grklpr3,watamura1,watamura2}, we briefly
describe those in \cite{watamura1,watamura2}. There is a left and
right action (``left'' and ``right'' `` regular representations''
$A^{L}$ and $A^{R}$) of $A$ on $A$: $b^{L}a = ba$ and $b^{R}a=ab, (b,
a \in A , b^{L,R}{\in}A^{L,R})$ with corresponding angular momentum
operators $L_{i}^{L}$ and $L_{i}^{R}$ and fuzzy coordinates
$x_{i}^{L}$ and $x_{i}^{R}$. $D$ and ${\gamma}$ are
\begin{eqnarray}
D &=&{\epsilon}_{ijk}{\sigma}_{i}x_{j}^{L}L_{k}^{R}, \nonumber\\
{\gamma}&=&-\frac{{\sigma}.\vec{L}^{R} -1/2}{l+1/2}. \label{fuzzyoprs}
\end{eqnarray}
Identifying $A^L$ as the representation of the fuzzy version of ${\cal
A}$, we have as before,
\begin{eqnarray}
\gamma b^L&=& b^L \gamma, \nonumber \\
{\gamma}D + D{\gamma}&=&0, \nonumber\\
{\gamma}^2&=&1, \nonumber\\
{\gamma}^{\dagger}&=&{\gamma}.
\end{eqnarray}
The carrier space of $A^{L, R}$, $D$ and ${\gamma}$ is $A^{2}$. When
$p^{(\pm N)}$ are also included, it gets expanded to $A^{2^{N+1}}$ as
$\vec{\tau}^{(i)}$ commute with $\vec{\sigma}$. Note that $ p^{(\pm
N)}$ commute with ${\gamma}$, as the $x$'s they contain are now being
identified with $x^{L}$'s.

We now construct a certain generalization of (\ref{wno}) for the fuzzy
sphere. It looks like (\ref{wno}), or rather the following expression:
\begin{eqnarray}
\pm N &=& -Tr_{\omega} \left( \frac{1}{|{\cal D}|^2}\Gamma \;{\cal
P}^{(\pm N)}\; 
[{\cal D}, {\cal P}^{(\pm N)}] \;[{\cal D}, {\cal P}^{(\pm N)}]\;
\right),\nonumber\\
|{\cal D}|&=&{\rm{Positive~ square~ root~ of}}~ {\cal
D}^{\dagger}{\cal D}, 
\end{eqnarray}
where ${\cal F}={\cal D}/|{\cal D}|$
\cite{connes,landi,coquereaux,vargra,mssv,varilly}. It is equivalent
to (\ref{wno}). It involves a Dixmier trace $Tr_{\omega}$ and
furthermore the inverse of $|{\cal D}|$.

But the massless Dirac operator on $A^{2^{N+1}}$ has zero modes and
$|D|$ has no inverse. An easy proof is as follows. We can write the
elements of $ A^{2^{N+1}} $ as rectangular matrices with entries
${\xi}_{{\lambda}j}{\in}A$ $({\lambda}=1,2,...,2^{N};j=1,2)$ where
$\lambda$ carries the action of ${\vec{\tau}}^{(i)}$'s and $j$ carries
the action of ${\vec{\sigma}}$. The dimensions of the subspaces
$U_\pm$ of $A^{2^{N+1}}$ with $\gamma =\pm 1$ are $[2(l \pm
1/2)+1][(2l+1)2^{N}]$. The first factor is the row dimension of
$U_\pm$ and is deduced from the fact that $(-\vec{L}^R +
\vec{\sigma}/2)^2$ has the definite values $(l+1/2)(l+3/2)$ and
$(l-1/2)(l+1/2)$ on $U_\pm$. The second factor is the column dimension
of $U_\pm$. $D$ anticommutes with $\gamma$. So if $D^{(+)}$ is the
restriction of $D$ to the domain $U_+$, $D^{(+)} = D|_{U_+} : U_+
\rightarrow U_-$, its index is ${\rm dim}\; U_+ - {\rm dim}\; U_- =
2[(2l+1)2^{N}]$. This is the minimum number of zero modes of $D$ in
$U_+$. Calculations \cite{watamura1,watamura2} show this to be the
exact number of zero modes, $D$ having no zero mode in $U_-$. In any
case, $D^{(+)}$ and so $D$ have no inverse. So we work instead with
the massive Dirac operator $D_m=D+m\gamma\;(m \neq 0)$ with the
strictly positive square $D^2_m=D^2+m^2$ and form the operator
\begin{equation}
f_m= \frac{D_m}{|D_m|}
\end{equation}
where
\begin{equation} 
|D_{m}|={\rm{Positive~ square~ root~ of}}~ D_m^{\dagger}D_m~, \quad
f_m^{\dagger}=f_m~, \quad f^2_m = {\bf1}.    
\end{equation} 
Consider $\frac{1-{\gamma}}{2}p^{(N)}f_m p^{(N)}\frac{1+{\gamma}}{2}$
where we pick $p^{(N)}$ and not $p^{(-N)}$ for specificity.  It
anticommutes with ${\gamma}$. Let ${\hat{V}}_{\pm} =p^{(N)}
U_{\pm}$. It then follows that the index of the operator
\begin{equation} 
{\hat{f}}^{(+)}_m = \frac{1-{\gamma}}{2} p^{(N)} f_m
p^{(N)}\frac{1+{\gamma}}{2} 
\end{equation}
(restricted to $\hat{V}_{+}$, such restrictions are hereafter to be
understood) is dimension of ${\hat{V}}_{+} $ $ - $ that of $
{\hat{V}}_{-} $ $ =2[2l+1+N]$. The index of its adjoint
\begin{equation} 
{\hat{f}}^{(+)\dagger}_m \equiv {\hat{f}}^{(-)}_m = \frac{1+{\gamma}}{2}
p^{(N)} f_m p^{(N)}\frac{1-{\gamma}}{2}
\end{equation} 
is $-2[2l+1+N]$.

We may try to associate the index of $ {\hat{f}}^{(+)}_m $ say with
the winding number $N$. But that will not be correct: this index is
not zero for $N=0$. The source of this unpleasant feature is also a
set of unwanted zero modes. Their presence can be established by
looking at ${\hat{f}}^{(\pm)}_m$ more closely. $ {\hat{f}}^{(\pm)}_m$
and $\gamma$ commute with ``total angular momentum'' $\vec{J} =
{\vec{L}}^{L} - {\vec{L}}^{R} + \sum_{i}\frac{\vec{\tau}^{(i)}}{2} +
\frac{\vec{\sigma}}{2}$ while $\gamma$ anticommutes with
${\hat{f}}^{(\pm)}_m$. So if an irreducible representation (IRR) of
$\vec{J}$ with $\vec{J}^2=j(j+1){\bf 1}$ occurs an odd number of times
in $\hat{V}_{+} + \hat{V}_{-}$, ${\hat{f}}^{(+)}_m +
{\hat{f}}^{(-)}_m$ must vanish on at least one of the $(2j+1)-$
dimensional eigenspaces. The remaining $(2j+1)-$ dimensional
eigenspaces can pair up so as to correspond to eigenvalues
${\pm}{\lambda}{\neq}0$ and get interchanged by $\gamma$. There are
two such $j$, both in $ \hat{V}_{+} $. They label IRR's with
multiplicity 1 and are its maximum and minimum $j^{(N)}=2l
+\frac{N+1}{2}$ and $\frac{N-1}{2}$. We can see that their eigenspaces
have ${\gamma}=+1$ as follows: the angular momentum value of
$\vec{L}^{L} + \sum_{i} {\frac{\vec{\tau}^{(i)}}{2}}$ in
$p^{(N)}A^{2^{N+1}}$ is $l+N/2$ so that the angular momentum value of
$-\vec{L}^{R}+\frac{\vec{\sigma}}{2}$ must be $l+1/2$ to attain the
$j-$ values $j^{(N)}$ and $\frac{N-1}{2}$. A further point is that
since $( 2j^{(N)} +1 ) + [2(\frac{N-1}{2})+1]$ is the index of
${\hat{f}}^{(+)}_m$ found earlier, we can conclude that there are no
other obligatory zero modes. Indeed every other $j$ labels IRR's of
multiplicity 2, one with ${\gamma}=+1$ and the other with $\gamma=-1$.

The zero modes for $j^{(N)}$ are unphysical as discussed by
Watamuras \cite{watamura1,watamura2}: there are no similar modes in
the continuum. If we can project them out, the index will shrink to $2
\frac{N-1}{2} +1 = N$, just what we want. So let ${\pi}^{(j^{(N)})}$
be the projection operator for $j^{(N)}$, constructed in the same
fashion as $p^{(N)}$. It commutes with $p^{(N)}$ since $p^{(N)}$
commutes with $ \vec{J}$. In fact, $p^{(N)}{\pi}^{(j^{(N)})} =
{\pi}^{(j^{(N)})}$ since if $j$ is maximum, then so is $k$. We thus
find that
\begin{equation} 
{\Pi}^{(N)}=p^{(N)} [ {\bf 1} - {\pi}^{(j^{(N)})} ]=p^{(N)} -
{\pi}^{(j^{(N)})}  
\end{equation}
is a projector. It commutes with $\gamma$ too. Let
\begin{eqnarray}
V_{\pm}&=&{\Pi}^{(N)}U_{\pm}, \nonumber\\
f^{(\pm)}_{m} &=& \frac{1{\mp}\gamma}{2} {\Pi}^{(N)} f_{m} {\Pi}^{(N)}
\frac{1{\pm}\gamma}{2} 
\end{eqnarray}
where ${f^{(+)}_m}^{\dagger}=f^{(-)}_m$. Then $f^{(+)}_m$ (restricted
to $V_{+}$) has the index $N$ we want.

The eigenvalues of $f^{(-,+)}_m \; f^{(+,-)}_m$ are the same (not
counting degeneracy) and for nonzero eigenvalues, the dimensions of
the corresponding eigenspaces are also identical. (We omit the
elementary proofs.) Therefore,
\begin{eqnarray} 
Tr\frac{1+\gamma}{2}{\Pi}^{(N)}[{\bf 1} - f^{(-)}_m
f^{(+)}_m]&-&\nonumber\\Tr\frac{1-\gamma}{2}{\Pi}^{(N)}[{\bf 1} -
f^{(+)}_m f^{(-)}_m]&=&\nonumber \\
Tr\frac{1+\gamma}{2}{\Pi}^{(N)} - Tr\frac{1-\gamma}{2}{\Pi}^{(N)} &=&
N\nonumber\\&=&\text{Index of}\quad f^{(+)}_m. \label{ncindex} 
\end{eqnarray} 

We want to be able to write (\ref{ncindex}) as a cyclic cocycle coming
from a Fredholm module
\cite{connes,landi,coquereaux,vargra,mssv,varilly}. The latter for us
is based on a representation $\Sigma$ of $A^{L}{\otimes}A^{R}$ on a
Hilbert space, and operators $F$ and $\epsilon$ with the following
properties:
\begin{eqnarray}
(i)\; F^{\dagger} &=& F, \quad F^2 = {\bf 1}. \\
(ii)\; \epsilon^{\dagger} &=& \epsilon, \quad \epsilon^2 = 1, \quad
\epsilon \Sigma(\alpha) = \Sigma(\alpha)\epsilon, \quad \epsilon F = -
F \epsilon 
\end{eqnarray}
where ${\alpha}{\in}A^{L}{\otimes}A^{R}$.  [This gives an {\it even}
Fredholm module, there need be no $\epsilon$ in an odd one.] We choose
for $\Sigma$ the representation
\begin{equation}
\Sigma:\alpha \rightarrow \Sigma(\alpha) = \left( \begin{array}{cc}
                                      \alpha & 0 \\
                                       0  & \alpha 
                                   \end{array} 
                             \right)
\end{equation}
on $A^{2^{N+1}} \oplus A^{2^{N+1}}$ and set 
\begin{equation}
F = \left( \begin{array}{cc}
             0  & f_m \\
             f_m & 0
           \end{array} \right), \quad \epsilon=\left( \begin{array}{cc}
                                                    {\bf 1} & 0 \\
                                                        0  & -{\bf 1}
                                                 \end{array} \right). 
\end{equation}
Introduce the projector
\begin{equation}
P^{(N)}=\left( \begin{array}{cc}
           \frac{1+\gamma}{2} {\Pi}^{(N)} & 0 \\
                    0  & \frac{1-\gamma}{2} {\Pi}^{(N)}
         \end{array} \right). \label{projector}
\end{equation}
Then
\begin{equation}
(P^{(N)}FP^{(N)})^2 = \left( \begin{array}{cc}
                   f^{(+)\dagger}_m f^{(+)}_m & 0 \\
                      0  & f^{(-)\dagger}_m f^{(-)}_m
                 \end{array}
          \right).
\end{equation}
Therefore,
\begin{equation}
\text{Index of } f^{(+)}_m = {\rm Tr}\; \epsilon \;[P^{(N)} -
(P^{(N)}FP^{(N)})^2]. \nonumber
\end{equation}
But since \cite{connes}
\begin{eqnarray}
P^{(N)}-(P^{(N)}FP^{(N)})^2 &=& -P^{(N)}[F, P^{(N)}]^2 P^{(N)}, \\
\text{Index of } f^{(+)}_m &=& N = -{\rm Tr}\; \epsilon P^{(N)}\;[F,
P^{(N)}]\;[F, P^{(N)}]. \label{dncindex}
\end{eqnarray}
This is the formulation of (\ref{ncindex}) we aimed at and is the
analogue of (\ref{wno}). It is worth remarking that we can replace
$\epsilon$ by $\left(\begin{array}{cc} {\gamma} & 0 \\ 0 & {\gamma}
\end{array} \right)$ here since ${\epsilon}P^{(N)} =
\left(\begin{array}{cc} {\gamma} & 0 \\ 0 & {\gamma} \end{array}
\right) P^{(N)}$. 

In $p^{(-N)}A^{2^{N+1}}$ as well, the unwanted zero modes correspond
to the top value $j^{(-N)}=2l -\frac{N-1}{2}$ of ``total angular
momentum ''. Once they are suppressed, the remaining obligatory zero
modes are readily shown to have the $j-$ value $\frac{N-1}{2}$,
multiplicity $N$ and $\gamma=-1$. Let ${\pi}^{(j^{(-N)})}$ be the
projector for the top angular momentum. Then it can be projected out
by replacing $p^{(-N)}$ by
\begin{equation}
{\Pi}^{(-N)}=p^{(-N)} [ {\bf 1} - {\pi}^{(j^{(-N)})}],
\end{equation}
but now $p^{(-N)}{\pi}^{j^{(-N)}}{\neq}{\pi}^{j^{(-N)}}$. Substituting
${\Pi}^{(-N)}$ for ${\Pi}^{(N)}$ in (\ref{projector}), we define
$P^{(-N)}$ and then by using (\ref{dncindex}) can associate $-N$ too with
an index.

There is the topic of fuzzy $\sigma$-fields yet to be discussed in
this section. We first note that in $u$ defined earlier, $a_{\mu}$ is
to be identified with $a_{\mu}^{L}$. Let us extend
$u^{{\tilde{\otimes}}N}$ and $g^{{\tilde{\otimes}}N}$ from $A^{2^{N}}$
to $A^{2^{N+1}} = A^{2^{N}}{\otimes}{\mathbb C}^2$ and ${\cal
A}^{2^{N+1}} = {\cal A}^{2^{N}}{\otimes}{\mathbb C}^2$ so that they
act as identity on the last ${\mathbb C}^2$'s. We also extend them
further to$ A^{2^{N+1}}{\oplus}A^{2^{N+1}}\equiv
A^{2^{N+1}}{\otimes}{\mathbb C}^2$ and ${\cal
A}^{2^{N+1}}{\oplus}{\cal A}^{2^{N+1}}\equiv {\cal
A}^{2^{N+1}}{\otimes}{\mathbb C}^2$ so that they act as identity on
these last ${\mathbb C}^2$'s. Define also $Q(u) =
u^{{\tilde{\otimes}}N} Q (u^{{\tilde{\otimes}}N})^{-1}$ for an
operator $Q ( = Q({\bf 1}))$ on $A^{2^{N+1}} {\oplus}
A^{2^{N+1}}$. The right hand side of (\ref{dncindex}) is invariant
under the substitution $P^{(N)}{\rightarrow}P^{(N)}(u)$ without
changing $F$. So $P^{(N)}(u)$ is a candidate for a fuzzy winding
number $N$ $\sigma$-field in the present context whereas previously it
was $p^{(N)}(u)$. But we must justify this candidacy by looking at the
continuum limit. In that limit, $u^{{\tilde{\otimes}}N} {\rightarrow}
g^{{\tilde{\otimes}}N}$, ${\pi}^{(j^{(N)})} {\rightarrow}
{\pi}^{(j^{(N)})}_{\infty}$ say and ${\Pi}^{(N)} {\rightarrow}
{\Pi}^{(N)}_{\infty}={\cal P}^{(N)} - {\pi}^{(j^{(N)})}_{\infty}$. The
stability group of ${\cal P}^{(N)}$ under conjugation by
$g^{{\tilde{\otimes}}N}$ is as before $U(1)$ at each $\vec{n}$. Now
${\pi}^{(j^{(N)})}$ projects out states where any one of $(\vec{L}^{L}
+ \frac{{\vec{\tau}}^{(i)}}{2})^2$, $(\vec{L}^{L} +
\frac{\vec{\sigma}}{2})^2$, $(\vec{L}^{L}-\vec{L}^{R})^2$ has the
maximum value. (Then any other pair of angular momenta also adds up to
maximum value as we saw in Section $3$.) So
${\vec{\tau}}^{(i)}. {\vec{x}^{L}}{\pi}^{(j^{(N)})} =
{\vec{\sigma}}.{\vec{x}^{L}}{\pi}^{(j^{(N)})}={\pi}^{(j^{(N)})}$,
${\vec{x}}^{L}. {\vec{x}}^{R}{\pi}^{(j^{(N)})}=-{\pi}^{(j^{(N)})}$. The
last condition is just a rule telling us that as
$l{\rightarrow}{\infty}$, $\vec{x}^{R}$ becomes $-\hat{n}$ on vectors
projected by ${\pi}^{(j^{(N)})}$. It will not show up in the continuum
projector. This establishes that ${\pi}^{(j^{(N)})}_{\infty}$ can be
identified with ${\cal P}^{(N+1)}=
(\prod_{i=1}^{N}\frac{1+\vec{\tau}^{(i)}.\hat{n}}{2}
)\frac{1+\vec{\sigma}.\hat{n}}{2}$ while of course
$\gamma{\rightarrow}\Gamma$. So ${\cal P}^{(N+1)}$ and
$\frac{1{\pm}\Gamma}{2}{\Pi}^{(N)}_{\infty}$ have $U(1)$ stability
groups, ${\Pi}^{(N)}_{\infty}(g)$ is a $\sigma$-field on $S^2$ and
$P^{(N)}(u)$ is a good choice for the fuzzy $\sigma$-field.

For winding number $-N$, we propose $P^{(-N)}(u)$ as the fuzzy
$\sigma$-field. We can check its validity also by going to the
continuum limit. As $l{\rightarrow}{\infty}$, $p^{(-N)}{\rightarrow}
{\cal P}^{(-N)}=\prod_{i=1}^{N}
\frac{1-{\vec{\tau}^{(i)}}.\hat{n}}{2}$ and has the $U(1)$ stability
group at each $\vec{n}$. Next consider the product
$p^{(-N)}{\pi}^{(j^{(-N)})}$. The presence of $p^{(-N)}$ allows us to
assume that $({\vec{K}}^{(N)})^2=k_{min}(k_{min}+1)$,
$k_{min}=l-N/2$. Also we can substitute for ${\pi}^{(j^{(-N)})}$ the
projector coupling ${\vec{K}^{(N)}}$ with $-\vec{L}^{R} +
\frac{\vec{\sigma}}{2}$ to give maximum angular momentum. This
projector for $l{\rightarrow}{\infty}$ gives the projector $\frac{1 +
{\vec{\sigma}}.\hat{n}}{2}$, the $N$ giving no contribution. (We also
get the condition ${\vec{x}}^R{\rightarrow}-\hat{n}$). So $p^{(-N)}
{\pi}^{(j^{(-N)})}$ as $l{\rightarrow}{\infty}$ can be identified with
${\cal P}^{(-N)}\frac{1+\vec{\sigma}.\hat{n}}{2}$ and that too has the
$U(1)$ stability group at each $\vec{n}$. This shows that
$P^{(-N)}(u)$ is a good fuzzy $\sigma$-field for winding number $-N$.

\section{\it Dynamics and Continuum Limit for Fuzzy $\sigma-$ Models}
The simplest action for the $O(3)$ nonlinear $\sigma$-model on
$S^2$ is
\begin{eqnarray}
&S&=\frac{\beta}{2}\int\frac{dcos{\theta}d{\phi}}{4{\pi}}({\cal
L}_i{\Phi}_a)(\vec{n})({\cal L}_i{\Phi}_a)(\vec{n}),\nonumber\\ 
&\sum_{a=1}^{3}&{\Phi}_a(\vec{n})^2=1,~{\beta}>0 \label{nlsm}
\end{eqnarray}
where $-i{\cal L}_i$ are the angular momentum operators on $S^2$. It
fulfills the important bound \cite{belpol}
\begin{equation}
S{\geq}{\beta}N
\label{bound}
\end{equation}
where $N({\geq}0)$ or$-N$ is as usual the winding number of the map
$\vec{\Phi}: S^2{\rightarrow}S^2$:
\begin{equation}
{\rm{~Winding ~number ~of}}~{\vec{\Phi}}=\frac{1}{2} \int_{S^2}
\frac{dcos{\theta}d{\phi}}{4{\pi}}
{\epsilon}_{ijk}n_i{\epsilon}_{abc}{\Phi}_a ({\cal L}_j{\Phi}_b)({\cal
L}_k{\Phi}_c). 
\end{equation}
This bound is obtained by integrating the inequality 
\begin{equation}
({\cal L}_i{\Phi}_a {\pm} {\epsilon}_{ijk}
n_j{\epsilon}_{abc}{\Phi}_b{\cal L}_k{\Phi}_c)^2{\geq}0  
\label{inequality}
\end{equation}
and is saturated if and only if 
\begin{equation}
{\cal L}_i{\Phi}_a {\pm} {\epsilon}_{ijk}
n_j{\epsilon}_{abc}{\Phi}_b{\cal L}_k{\Phi}_c = 0 
\label{saturation}
\end{equation}
for one choice of sign. The solutions of (\ref{saturation}) can be
thought of as two dimensional instantons \cite{belpol}.

We now propose a fuzzy $\sigma$-action using these properties of
$S$ as our guide. Consider the inequality
\begin{equation}
([F,P(u)]\frac{1{\pm}{\epsilon}}{2}P(u))^{\dagger}
( [F,P(u)]\frac{1{\pm}{\epsilon}}{2}P(u)){\geq}0 
\end{equation}
where $P(u)$ can be $P^{(N)}(u)$ or $P^{(-N)}(u)$ and $Q{\geq}0$ here
means that $Q$ is a nonnegative operator. This is the analogue of
(\ref{inequality}). Taking trace, we get the analogue of (\ref{bound}),
\begin{equation}
s_F{\equiv}Tr P(u) [F,P(u)][F,P(u)]{\geq}N.
\label{dinequality}   
\end{equation}
The bound is saturated if and only if 
\begin{equation}
[F,P(u)]\frac{1{\pm}{\epsilon}}{2}P(u)=0 
\label{dsaturation}
\end{equation}
for one choice of sign, just like in (\ref{saturation}). All this
suggests the novel fuzzy $\sigma$-action
\begin{equation}
S_F={\beta}_Fs_{F}.  
\end{equation}

Qualitative remarks about the approach to continuum of $S_F$ will now
be made. The first is that $\beta_F$ and $m$ must be scaled as
$l{\rightarrow}{\infty}$. As regards the scaling of $\beta_F$, we
conjecture that (\ref{dsaturation}) has no solution for finite $l$ and
that (\ref{dinequality}) is a strict inequality. Choose
\begin{equation}
{\Lambda}(l)= \frac{1}{N}{\times}(\text{Minimum of}\; s_F)
\end{equation}
so that $\frac{s_F}{\Lambda(l)}$ is $N$ at minimum. Then we suggest
that we should set
\begin{equation}
{\beta}_F=\frac{\beta}{\Lambda(l)}.
\end{equation}
It is our conjecture too that $\Lambda (l)$ diverges as
$l{\rightarrow}{\infty}$ in such a way that (upto factors)
\begin{eqnarray}
S_F{\rightarrow}S_{\infty} &=& {\beta} Tr_{\omega} P_{\infty}(g)
[{\cal F}, P_{\infty}(g)][{\cal F},P_{\infty}(g)],\nonumber\\  
{\cal F} &=& \left(\begin{array}{cc}
                  0  &  {\cal D}/|{\cal D}| \\
                  {\cal D}/|{\cal D}|   &  0
               \end{array}
          \right), \nonumber \\
g &=& \lim_{l{\rightarrow}{\infty}}u,\nonumber\\ 
P_{\infty}(g) &=& \lim_{l{\rightarrow}{\infty}}P(u)
\end{eqnarray}
where we have let $m$ become zero as ${\cal D}$ has no zero mode. An
alternative form of $S_{\infty}$ is 
{\small 
\begin{equation}
S_{\infty} = {\beta}\int\frac{d\cos{\theta}d{\phi}}{4{\pi}} Tr 
   P_{\infty}(g) \left[\left(\begin{array}{cc} 
                  0  &  {\cal D} \\
                  {\cal D}   &  0
                \end{array}
          \right)\!, P_{\infty}(g)\right]\!\!\left[\left(\begin{array}{cc}
                  0  &  {\cal D} \\
                  {\cal D}   &  0
                \end{array}
          \right)\!, P_{\infty}(g)\right]
\end{equation}
}
where the trace $Tr$ is only over the internal indices. 

We now argue that $P(u)$ itself must be corrected by cutting off all
high angular momenta (and not just the top one) while passing to
continuum. Thus it was mentioned before that state vectors with top
``total'' angular momentum $j^{({\pm}N)}$ are unphysical. Their
characteristic feature is their divergence as $l \rightarrow
\infty$. That means that once normalized these vectors become weakly
zero in the continuum limit. In fact any sequence of vectors with a
linearly divergent $j$ as $l{\rightarrow}{\infty}$ is unphysical. Such
$j$ contribute eigenvalues to the Dirac operator which are nonexistent
in the continuum, as one can verify using the results of
\cite{watamura1,watamura2} for $N=0$: the spectrum of $D$ then is
${\pm}(j+\frac{1}{2})[ 1 + (1 - (j + \frac{1}{2})^2)/(4l(l+1))]^{1/2}$
while that of ${\cal D}$ is ${\pm}(j+\frac{1}{2})$, $j$ being total
angular momentum. The corresponding eigenvectors too if normalized are
weakly zero in the $l \rightarrow\infty$ limit. It seems necessary
therefore to eliminate them in a suitable sense during the passage to
the limit.

One way to do so may be to use a double limit which we now
describe. Let ${\pi}^{(J)}$ be the projection operator for all states
with $j{\geq}J$. Let us define
\begin{eqnarray}
P^{({\pm}N)(J)}&=&
 \left(\begin{array}{cc}
                  \frac{1+{\gamma}}{2}p^{(\pm N)}({\bf 1} -
 {\pi}^{(J)})  &  0 \\ 
                  0   &  \frac{1-{\gamma}}{2}p^{(\pm N)}({\bf 1} -
 {\pi}^{(J)}) 
                \end{array}
          \right), \nonumber\\
P^{{({\pm}N)}(J)}(u) &=& u^{{\tilde{\otimes}}N} P^{{({\pm}N)}(J)}
 [u^{{\tilde{\otimes}}N}]^{-1}. 
\end{eqnarray}  

We then consider the fuzzy $\sigma$-model with $P^{({\pm}N)(J)}(u)$
replacing $P^{({\pm}N)}(u)$ ${\equiv} P^{({\pm}N)(j^{(N)})}(u)$ and
thereby cutting off angular momenta ${\geq}J$. That would not affect
index theory arguments so long as $J>\frac{N-1}{2}$ as the important
zero modes will then be left intact. We are thus led to the cut-off
action
\begin{eqnarray}
S_F^{(J)} &=& \frac{\beta}{{\Lambda}^{(J)}(l)}s_F^{(J)},\nonumber\\
s_F^{(J)} &=& TrP^{({\pm}N)(J)}(u) [F,P^{({\pm}N)(J)}(u)]
[F,P^{({\pm}N)(J)}(u)],\nonumber\\ 
{\Lambda}^{(J)}(l) &=& {\frac{{\rm{~Minimum ~of}} ~s_F^{(J)}}{N}},
\end{eqnarray}
and the following suggestion: A good way to define the continuum
partition function is to let $l$ and $J{\rightarrow}{\infty}$ in that
order in the partition function of $S_F^{(J)}$. Thus we propose the
continuum partition function
\begin{equation}
Z=\lim_{J{\rightarrow}{\infty}} \lim_{l{\rightarrow}{\infty}}\int
d{\mu}~ exp (- S_F^{(J)}), 
\end{equation} 
$d{\mu}$ denoting the functional measure. The inner limit recovers the
continuum where the contributions of vectors with divergent $J$ should
not matter, for this reason this method may eliminate the influence of
unwanted modes from $Z$. Perhaps an equivalent limiting procedure
would be let $l,J$ ${\rightarrow}{\infty}$ with $J/l{\rightarrow}0$.

Taking the limit $ l{\rightarrow}{\infty} $ with fixed $J$ is
compatible with the continuum description of the $\sigma$-field. In
that limit, $p^{({\pm}N)}$ becomes ${\cal P}^{({\pm}N)}$.  Next
consider the vectors projected by $p^{({\pm}N)}[1-{\pi}^{(J)}]$. The
effect of the last factor on the projected vectors is as follows: For
$\gamma=1$ say, we must combine the angular momentum value
$l{\pm}\frac{N}{2}$ of $\vec{K}^{(N)}$ with the value $l+1/2$ of
$-\vec{L}^{R} + \frac{\vec{\sigma}}{2}$ to produce an allowed value
$j<J$ of any such vector. So $[\vec{K}^{(N)} + (-\vec{L}^{R} +
\frac{\vec{\sigma}}{2})]^2 = j(j+1)$, $(\vec{K}^{(N)})^2 =
(l{\pm}\frac{N}{2})(l{\pm}\frac{N}{2} + 1)$ and $(-\vec{L}^{R} +
\frac{\vec{\sigma}}{2})^2 = (l+\frac{1}{2})(l+\frac{3}{2})$. Letting
$l{\rightarrow}{\infty}$, we find that
$\vec{x}^L.\vec{x}^R{\rightarrow}1$ due to the factor
$[1-{\pi}^{(J)}]$, where we have used the fact that
$\frac{\vec{\tau}^{(i)}}{l}$ and $\frac{\vec{\sigma}}{l}
{\rightarrow}0$ as $l{\rightarrow}{\infty}$. But this is just a rule
instructing us to set $\vec{x}^R=\hat{n}$ for large $l$ for these
vectors, and will not show up in the continuum projector. The
$\gamma=-1$ case is no different in the continuum limit. Thus for
$l{\rightarrow}{\infty}$, $p^{({\pm}N)}(u)[1-{\pi}^{(J)}(u)]$ can be
interpreted as ${\cal P}^{({\pm}N)}(g) = g^{{\hat{\otimes}}N}{\cal P}
^{({\pm}N)}[g^{{\hat{\otimes}}N}]^{-1}$, the continuum
$\sigma$-fields.

Let   
\begin{equation}
 P^{({\pm} N)(J)}_{\infty}(g) = \lim_{l{\rightarrow}{\infty}}
P^{({\pm}N)(J)}(u) = \left(\begin{array}{cc}
                  \frac{1+{\Gamma}}{2}{\cal P}^{({\pm}N)}(g)  &  0 \\
                  0   & \frac{1-{\Gamma}}{2}{\cal P}^{({\pm}N)}(g) 
                \end{array}
          \right).
\end{equation}
Then the naive $l{\rightarrow}{\infty}$, $m{\rightarrow}0$ limit of
$S_F^{(J)}$ is expected to be (upto factors)
\begin{equation}
{\cal S}_{\infty} = {\beta}Tr_{\omega} P^{({\pm}N)(J)}_{\infty}(g) [{\cal
F},P^{({\pm}N)(J)}_{\infty}(g)][{\cal F},P^{({\pm}N)(J)}_{\infty}(g)]
\end{equation}
which can be simplified to 
\begin{equation}
{\cal S}_{\infty} = {\beta}\int\frac{dcos{\theta}d{\phi}}{4{\pi}}
Tr{\cal P}^{{(\pm)N}}(g)[{\cal D},{\cal P}^{{(\pm)N}}(g)][{\cal
D},{\cal P}^{{(\pm)N}}(g)].  
\end{equation}
It seems to correspond to (\ref{nlsm}).       
       
\section{\it Remarks}
\begin{itemize}
\item[-]The Dirac operators in (\ref{contoprs}) and (\ref{fuzzyoprs})
differ from those of \cite{watamura1,watamura2} by unitary
transformations generated by ${\Gamma}$ and ${\gamma}$.
 
\item[-]Nonabelian monopoles such as the elementary $U(2)$ monopoles
and their fuzzy versions can very likely be accommodated in our
approach using different projectors.

\item[-]In the same manner, there seems to be no big barrier to
studying the case of Grassmannians $G_{n,k}(\mathbb C) =
U(n+k)/[U(n){\times}U(k)]$ as target spaces in ${\sigma}$-models as
they are orbits of rank $n$ projectors under $U(n+k)$. We can also
imagine treating other target spaces by considering orbits under
subgroups of $U(n+k)$, the previous choice of $\{ g^{\tilde{\otimes}
N}(\vec{n}) \} {\subset} U(2N)$ being an example.

\item[-]We have managed to generalize the approach here to fuzzy
manifolds like fuzzy ${\mathbb C}{\mathbb P}^{2}$ (\cite{grostr}) or
more generally to fuzzy versions of orbits of simple Lie groups in the
adjoint representation. We will report on this work elsewhere.

\item[-]Further discussion of fuzzy quantum physics of monopoles and
solitons is needed to better reveal the implications of fuzzy quantum
physics in its topological aspects.

\item[-]The fuzzy Dirac operator $D'$ used in
\cite{gropre,grklpr1,grklpr2,grklpr3} gives a much better
approximation to the spectrum of the continuum Dirac operator. Its
eigenvalue for the top angular momentum state vector is largest in
modulus, recedes to infinity with $l$ and is not a zero mode as in the
case of $D$. The contribution of this vector therefore tends to be
suppressed in functional integrals. In \cite{baidgo}, a chirality
operator $\gamma'$ for $D'$ (with the correct continuum limit) has
been constructed after projecting out this vector. The contents of the
present paper can be easily recast using $D'$ and
$\gamma'$. Ref.\cite{baidva} discusses $\theta$-states and chiral
anomalies in gauge theories of fuzzy physics using these new
operators.

\end{itemize}

\bigskip

During this work we were fortunate to receive help and advice from
colleagues and friends like T. R. Govindarajan, Giorgio Immirzi,
C. Klim\v{c}\'{\i}k, Gianni Landi, Fedele Lizzi, Xavier Martin, Denjoe
O'Connor, Paulo Teotonio-Sobrinho and J. C. Varilly. Fedele was
especially helpful with detailed comments and pointed out an error in
an earlier version as also the connection of our work to that in
ref.\cite{grklpr1}. We thank them. A.P.B. also warmly thanks Fedele
Lizzi and Beppe Marmo for their exceptionally friendly hospitality and
for arranging support by INFN at Dipartimento di Scienze Fisiche,
Universit\`a di Napoli, while this work was being completed.

This work was supported in part by the DOE under contract number
DE-FG02-85ER40231. 
   
\bibliographystyle{unsrt}

\end{document}